\documentclass[12pt]{article}
\usepackage{graphicx}
\usepackage{xspace}
\def\pbnr{}
\def\speaker{Antimo Palano}
\def\onbehalfof{the LHCb Collaboration}
\def\title{Open Charm Spectroscopy and Mass Measurements in LHCb}
\def\affiliation{INFN and University of Bari, Italy}
\def\support{}

\newcommand\pubnumber{\pbnr}
\newcommand\pubdate{\today}
\def\Title#1{\begin{center} {\Large #1 } \end{center}}
\def\Author#1{\begin{center}{ \sc #1} \end{center}}

\newcommand{\OnBehalf}[1]{\sbox0{#1}\ifdim\wd0=0pt
        {}
	\else
	{\\on behalf of #1}
	\fi}
\newcommand{\SupportedBy}[1]{\sbox0{#1}\ifdim\wd0=0pt
        {}
	\else
	{\footnote{#1}}
	\fi}
\def\Address#1{\begin{center}{ \it #1} \end{center}}

\newcommand\pubblock{\includegraphics[width=5cm]{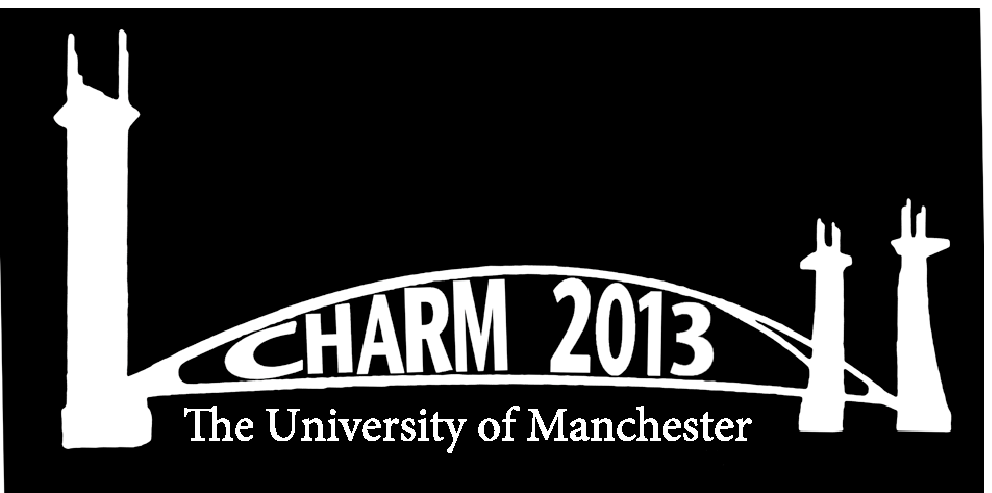}\hfill{\begin{tabular}{l} \pubnumber\\
         \pubdate  \end{tabular}}}
\newenvironment{Abstract}{\begin{quotation}  }{\end{quotation}}
\newenvironment{Presented}{\begin{quotation} \begin{center} 
             PRESENTED AT\end{center}\bigskip 
      \begin{center}\begin{large}}{\end{large}\end{center} \end{quotation}}
\def\Acknowledgements{\bigskip  \bigskip \begin{center} \begin{large}
             \bf ACKNOWLEDGEMENTS \end{large}\end{center}}
\def\venue{The 6$^{th}$ International Workshop on Charm Physics\\
(CHARM 2013)\\
Manchester, UK,  31 August -- 4 September, 2013}

\def\beq{\begin{equation}}
\def\eeq#1{\label{#1}\end{equation}}
\def\eeqn{\end{equation}}

\def\beqa{\begin{eqnarray}}
\def\eeqa#1{\label{#1}\end{eqnarray}}
\def\eeqan{\end{eqnarray}}

\let\bar=\overbar

\def\D{{\cal D}}

\def\Dslash{\not{\hbox{\kern-4pt $D$}}}
\def\dslash{\not{\hbox{\kern-2pt $\del$}}}

\def\msb{{\bar{\ssstyle M \kern -1pt S}}}

\def\lhcb {\mbox{LHCb}\xspace}
\def\ux85 {\mbox{UX85}\xspace}

\def\babar  {\mbox{BaBar}\xspace}
\def\belle  {\mbox{Belle}\xspace}

 \def\Ppi         {\ensuremath{\uppi}\xspace}

 \def\Pphi        {\ensuremath{\upphi}\xspace}

 \def\PDelta      {\ensuremath{\Delta}\xspace}                 
 \def\PXi      {\ensuremath{\Xi}\xspace}                 
 \def\PLambda      {\ensuremath{\Lambda}\xspace}                 
 \def\PSigma      {\ensuremath{\Sigma}\xspace}                 
 \def\POmega      {\ensuremath{\Omega}\xspace}                 
 \def\PUpsilon      {\ensuremath{\Upsilon}\xspace}                 
 
 \def\PB      {\ensuremath{\mathrm{B}}\xspace}                 
                  
 \def\PD      {\ensuremath{\mathrm{D}}\xspace}

 \def\PK      {\ensuremath{\mathrm{K}}\xspace}

 \def\Pi      {\ensuremath{\mathrm{i}}\xspace}

 \def\Ppi         {\ensuremath{\pi}\xspace}

 \def\Pphi        {\ensuremath{\phi}\xspace}

 \mathchardef\PDelta="7101
 \mathchardef\PXi="7104
 \mathchardef\PLambda="7103
 \mathchardef\PSigma="7106
 \mathchardef\POmega="710A
 \mathchardef\PUpsilon="7107
                  
 \def\PB      {\ensuremath{B}\xspace}                 
                  
 \def\PD      {\ensuremath{D}\xspace}

 \def\PK      {\ensuremath{K}\xspace}

 \def\Pi      {\ensuremath{i}\xspace}

\def\pion  {\ensuremath{\Ppi}\xspace}

\def\pip   {\ensuremath{\pion^+}\xspace}
\def\pim   {\ensuremath{\pion^-}\xspace}

\def\kaon  {\ensuremath{\PK}\xspace}
  \def\Kbar  {\kern 0.2em\overline{\kern -0.2em \PK}{}\xspace}

\def\Kz    {\ensuremath{\kaon^0}\xspace}
\def\Kzb   {\ensuremath{\Kbar^0}\xspace}
\def\KzKzb {\ensuremath{\Kz \kern -0.16em \Kzb}\xspace}
\def\Kp    {\ensuremath{\kaon^+}\xspace}
\def\Km    {\ensuremath{\kaon^-}\xspace}

\def\KpKm  {\ensuremath{\Kp \kern -0.16em \Km}\xspace}

  \def\Dbar    {\kern 0.2em\overline{\kern -0.2em \PD}{}\xspace}
\def\D       {\ensuremath{\PD}\xspace}

\def\Dz      {\ensuremath{\D^0}\xspace}
\def\Dzb     {\ensuremath{\Dbar^0}\xspace}
\def\DzDzb   {\ensuremath{\Dz {\kern -0.16em \Dzb}}\xspace}
\def\Dp      {\ensuremath{\D^+}\xspace}
\def\Dm      {\ensuremath{\D^-}\xspace}

\def\DpDm    {\ensuremath{\Dp {\kern -0.16em \Dm}}\xspace}
\def\Dstar   {\ensuremath{\D^*}\xspace}

\def\Dstarp  {\ensuremath{\D^{*+}}\xspace}

\def\B       {\ensuremath{\PB}\xspace}
  \def\Bbar    {\kern 0.18em\overline{\kern -0.18em \PB}{}\xspace}

  \def\Y#1S{\ensuremath{\PUpsilon{(#1S)}}\xspace}

\def\Lbar {\ensuremath{\kern 0.1em\overline{\kern -0.1em\Lambda\kern -0.05em}\kern 0.05em{}}\xspace}

\def\to                 {\ensuremath{\rightarrow}\xspace}

\def\AT#1     {\ensuremath{A_{\mathrm{T}}^{#1}}\xspace}           

\def\C#1      {\ensuremath{\mathcal{C}_{#1}}\xspace}                       
\def\Cp#1     {\ensuremath{\mathcal{C}_{#1}^{'}}\xspace}                    
\def\Ceff#1   {\ensuremath{\mathcal{C}_{#1}^{\mathrm{(eff)}}}\xspace}        
\def\Cpeff#1  {\ensuremath{\mathcal{C}_{#1}^{'\mathrm{(eff)}}}\xspace}       
\def\Ope#1    {\ensuremath{\mathcal{O}_{#1}}\xspace}                       
\def\Opep#1   {\ensuremath{\mathcal{O}_{#1}^{'}}\xspace}                    

\newcommand{\tev}{\ensuremath{\mathrm{\,Te\kern -0.1em V}}\xspace}
\newcommand{\gev}{\ensuremath{\mathrm{\,Ge\kern -0.1em V}}\xspace}
\newcommand{\mev}{\ensuremath{\mathrm{\,Me\kern -0.1em V}}\xspace}
\newcommand{\kev}{\ensuremath{\mathrm{\,ke\kern -0.1em V}}\xspace}
\newcommand{\ev}{\ensuremath{\mathrm{\,e\kern -0.1em V}}\xspace}
\newcommand{\gevc}{\ensuremath{{\mathrm{\,Ge\kern -0.1em V\!/}c}}\xspace}
\newcommand{\mevc}{\ensuremath{{\mathrm{\,Me\kern -0.1em V\!/}c}}\xspace}
\newcommand{\gevcc}{\ensuremath{{\mathrm{\,Ge\kern -0.1em V\!/}c^2}}\xspace}
\newcommand{\gevgevcccc}{\ensuremath{{\mathrm{\,Ge\kern -0.1em V^2\!/}c^4}}\xspace}
\newcommand{\mevcc}{\ensuremath{{\mathrm{\,Me\kern -0.1em V\!/}c^2}}\xspace}

\def\invfb   {\ensuremath{\mbox{\,fb}^{-1}}\xspace}

\def\gsim{{~\raise.15em\hbox{$>$}\kern-.85em
          \lower.35em\hbox{$\sim$}~}\xspace}
\def\lsim{{~\raise.15em\hbox{$<$}\kern-.85em
          \lower.35em\hbox{$\sim$}~}\xspace}

\def\pt         {\mbox{$p_{\rm T}$}\xspace}

\def\tell1  {TELL1\xspace}
\def\ukl1   {UKL1\xspace}

\def\DJ     {\ensuremath{\D_J}\xspace}
\def\DstarPi {\ensuremath{\Dstarp \pim}\xspace}
\def\DPi {\ensuremath{\Dp \pim}\xspace}
\def\DzPi {\ensuremath{\Dz \pip}\xspace}
\def\DTwentyFourThirty {\ensuremath{{D}^\prime_1(2430)}\xspace}
\def\DTwentyFourTwenty {\ensuremath{D_1(2420)}\xspace}
\def\DTwentyFourTwentyNeutral {\ensuremath{ D_1(2420)^0}\xspace}
\def\DTwentyFourTwentyCharged {\ensuremath{ D_1(2420)^+}\xspace}
\def\DTwentyFourSixty {\ensuremath{D^*_2(2460)}\xspace}
\def\DTwentyFourSixtyNeutral {\ensuremath{ {D}^*_2(2460)^0}\xspace}
\def\DTwentyFourSixtyCharged {\ensuremath{ {D}^*_2(2460)^+}\xspace}
\def\DTwentyFourHundred {\ensuremath{D^*_0(2400)}\xspace}

\def\DTwentyFiveFiftyNeutral {\ensuremath{D_{J}(2580)^0}\xspace}

\def\DTwentySixHundredNeutral {\ensuremath{D^*_{J}(2650)^0}\xspace}

\def\DTwentySevenFiftyNeutral {\ensuremath{D_{J}(2740)^0}\xspace}

\def\DTwentySevenSixtyNeutral {\ensuremath{D^*_{J}(2760)^0}\xspace}
\def\DTwentySevenSixtyCharged {\ensuremath{D^*_{J}(2760)^+}\xspace}

\def\DThreeNeutral {\ensuremath{D^*_{J}(3000)^0}\xspace}
\def\DThreeCharged {\ensuremath{D^*_{J}(3000)^+}\xspace}
\def\DThreeU {\ensuremath{D_{J}(3000)^0}\xspace}
\def\invfb   {\ensuremath{\mbox{\,fb}^{-1}}\xspace}

\def\mthetah     {\mbox{$\theta_{\rm H}$}\xspace}

\def\cthetah     {\mbox{$\cos\theta_{\rm H}$}\xspace}
\def\DTwentyFiveFiftyNeutralb {\ensuremath{{D}(2550)^0}\xspace}

\def\DTwentySixHundredNeutralb {\ensuremath{{D^*}(2600)^0}\xspace}
\def\DTwentySixHundredChargedb {\ensuremath{{D^*}(2600)^+}\xspace}
\def\DTwentySevenFiftyNeutralb {\ensuremath{{D}(2750)^0}\xspace}

\def\DTwentySevenSixtyNeutralb {\ensuremath{{D^*}(2760)^0}\xspace}
\def\DTwentySevenSixtyChargedb {\ensuremath{{D^*}(2760)^+}\xspace}

\begin{document}
\begin{titlepage}
\pubblock

\vfill
\Title{\title}
\vfill
\Author{\speaker\SupportedBy{\support}\OnBehalf{\onbehalfof}}
\Address{\affiliation}
\vfill
\begin{Abstract}
A study of \DPi, \DzPi and \DstarPi final states is performed using $pp$ collision data, corresponding to an integrated luminosity of 1.0\invfb, 
collected at a centre-of-mass energy of $7\tev$ with the \lhcb detector. 
The \DTwentyFourTwentyNeutral resonance is observed in the  \DstarPi final state and the \DTwentyFourSixty resonance is observed in the 
\DPi, \DzPi and \DstarPi final states. For both resonances, their properties and spin-parity assignments are obtained.
In addition, two natural parity and two unnatural parity resonances are observed in the mass region between 2500 and 2800~\mev.
Further structures in the region around 3000 \mev are observed in all the \DstarPi, \DPi and \DzPi final states.
Using three- and four-body decays of $D$ mesons produced in semileptonic
$b$-hadron decays, precision measurements of $D$ meson mass differences are
made together with a measurement of the $D^{0}$ mass.

\end{Abstract}
\vfill
\begin{Presented}
\venue
\end{Presented}
\vfill
\end{titlepage}
\def\thefootnote{\fnsymbol{footnote}}
\setcounter{footnote}{0}
%

\section{Introduction}
Charm meson spectroscopy provides a powerful test of the quark model predictions of the Standard Model.
Many charm meson states, predicted in the 1980s~\cite{Godfrey:1985xj}, have not yet been observed experimentally. 
The $J^P$ states having $P=(-1)^J$ and therefore $J^P=0^+,1^-,2^+,...$ are called 
natural parity states and are labelled as $D^*$, while unnatural parity indicates the series $J^P=0^-,1^+,2^-,...$. 
 Apart from the ground states ($D,D^*$), only two of the 1P states, \DTwentyFourTwenty and \DTwentyFourSixty, are experimentally well established since they have relatively narrow widths ($\sim$30\mev).~\footnote{We work in units where $c = 1$.} In contrast, the broad $L=1$ states, \DTwentyFourHundred and \DTwentyFourThirty, have been established by the \belle and \babar \ experiments in exclusive \B decays~\cite{Abe:2003zm,Aubert:2009wg}. A search for excited charmed mesons, labelled \DJ, has been performed by \babar~\cite{delAmoSanchez:2010vq}. They observe four signals, labelled \DTwentyFiveFiftyNeutralb, \DTwentySixHundredNeutralb, \DTwentySevenFiftyNeutralb and \DTwentySevenSixtyNeutralb, and the 
isospin partners \DTwentySixHundredChargedb and \DTwentySevenSixtyChargedb. 

This study~\cite{Aaij:2013sza} reports a search for \DJ mesons in a data sample, corresponding to an integrated luminosity of 1.0\invfb, of $pp$ collisions collected at a centre-of-mass energy of $7\tev$ with the \lhcb detector. 

\section{Event selection}
\label{sec:Selection}

The search for \DJ mesons is performed using the inclusive reactions
\begin{equation}
pp \to \Dp \pim X, \ pp \to \Dz \pip X , \ pp \to \Dstarp \pim X,
\end{equation}
where $X$ represents a system composed of any collection of charged and neutral particles~\footnote{Throughout the paper use of charge-conjugate decay modes is implied.}.

The charmed mesons in the final state are reconstructed in the decay modes ${\mbox \Dp \to K^-\pi^+\pi^+}$, $\Dz \to \Km \pip$ and $\Dstarp \to \Dz \pip$.
Charged tracks are required to have good track fit quality, momentum $p>3\gev$ and $\pt>250\mev$. 
These conditions are relaxed to lower limits for the pion originating directly from the \Dstarp decay. 
The cosine of the angle between the momentum of the $D$ meson candidate and its 
direction, defined by the positions of the primary vertex and the meson decay vertex, is required to be larger than 0.99999. This ensures that the $D$ meson candidates are produced at the 
primary vertex and reduces the contribution from particles originating from $b$-hadron decays.
The purity of the charmed meson candidates is enhanced by requiring the decay products to be identified by the RICH detectors.

The reconstructed  $D^+$, $D^0$ and $\Dstarp$ candidates are combined with all the right-sign charged pions in the event.
Each of the  $D^+ \pim$, the $D^0 \pip$, and the $\Dstarp \pim$ candidates are fitted to a common vertex with $\chi^2/{\rm ndf}<8$, where ndf is the number of degrees of freedom. 

In order to reduce combinatorial background, the cosine of the angle between the momentum direction of the charged pion in the $D^{(*)}\pi^{\pm}$ rest frame and the momentum direction of the $D^{(*)}\pi^{\pm}$ system in the laboratory frame is required to be greater than zero. 
It is also required that the $D^{(*)}$ and the $\pi^{\pm}$ point to the same primary vertex. 

\section{Mass spectra}
\label{sec:mass}
\begin{figure}[htb]
\centering
\includegraphics[height=1.5in]{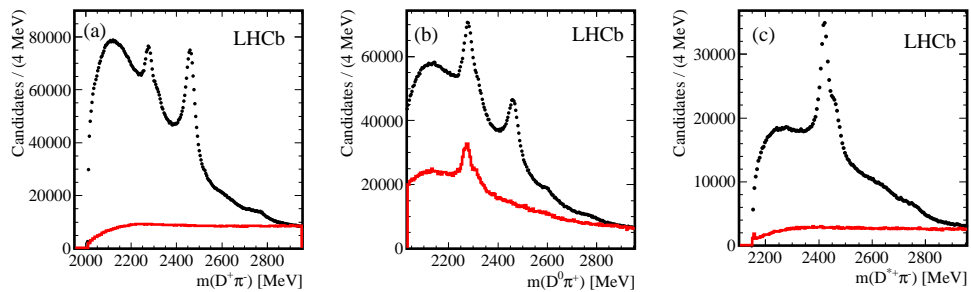}
  \caption{
    \small Invariant mass distribution for (a) $\Dp \pim$,  (b) $\Dz \pip$ and (c) $\Dstarp \pim$ candidates (points). The full line histograms (in red) show the wrong-sign mass spectra for (a) $\Dp \pip$,  (b) $\Dz \pim$ and (c) $\Dstarp \pip$ normalized to the same yield at high $D^{(*)}\pi$ masses.
    }
  \label{fig:fig3}
\end{figure}

The $\Dp \pim$, $\Dz \pip$ and $\Dstarp \pim$ mass spectra are shown in Fig.~\ref{fig:fig3}.
A further reduction of the combinatorial background is achieved by performing an optimization of the signal significance and purity as a function of  
$p_{\rm T}$ of the $D^{(*)}\pi^{\pm}$ system using the well known \DTwentyFourTwenty and \DTwentyFourSixty resonances.~\footnote{We use the generic notation $D$ to indicate both neutral and charged $D$ mesons.} 
After the optimization 7.9$\times 10^6$, 7.5$\times 10^6$ and 2.1$\times 10^6$ $\Dp \pim$, $\Dz \pip$ and $\Dstarp \pim$ candidates are obtained.

We analyze, for comparison and using the same selections, the wrong-sign $\Dp \pip$, $\Dz \pim$ and $\Dstarp \pip$ combinations which are also shown in Fig.~\ref{fig:fig3}.
 
The $\Dp \pim$ mass spectrum, Fig.~\ref{fig:fig3}(a), shows a double peak structure around 2300~\mev due to cross-feed from the decay
\begin{equation}
\DTwentyFourTwentyNeutral \ {\rm or} \ \DTwentyFourSixtyNeutral \to \pi^- D^{*+} (\to D^+ \pi^0/\gamma )  \ (32.3\%),
\label{dp_pim}
\end{equation}
where the $\pi^0/\gamma$ is not reconstructed; the last number, in parentheses, indicates the branching fraction of $D^{*+} \to  D^+ \pi^0/\gamma$ decays.
We observe a strong \DTwentyFourSixtyNeutral signal and weak structures around 2600 and 2750~\mev. The wrong-sign $\Dp \pip$ mass spectrum does not show any structure.

The $\Dz \pip$ mass spectrum, Fig.~\ref{fig:fig3}(b), shows an enhanced double peak structure around 2300~\mev
due to cross-feed from the decays
\begin{equation}
\DTwentyFourTwentyCharged  \ {\rm or} \ \DTwentyFourSixtyCharged \begin{array}{l} \to \pi^+ D^{*0} \begin{array}{l}  \\ ( \to  D^0 \pi^0)  \ (61.9\%) \\ (\to  D^0 \gamma)  \ (38.1\%) \ . \end{array} \end{array} 
\label{dz_pip}
\end{equation}
The \DTwentyFourSixtyCharged signal and weak structures around 2600 and 2750~\mev are observed. 
In comparison, the wrong-sign $\Dz \pim$ mass spectrum does 
show the presence of structures in the 2300~\mev mass region, similar to those observed in the $\Dz \pip$ mass spectrum.
These structures are due to cross-feed from the decay
\begin{equation}
\DTwentyFourTwentyNeutral \ {\rm or} \ \DTwentyFourSixtyNeutral \to \pi^- D^{*+} ( \to   D^0 \pi^+)  \ (67.7\%) \ .
\label{dz_pim}
\end{equation}

The $\Dstarp \pim$ mass spectrum, Fig.~\ref{fig:fig3}(c), is dominated by the presence of the \DTwentyFourTwentyNeutral and \DTwentyFourSixtyNeutral signals. At higher mass, complex broad structures are evident in the mass region between 2500 and 2800~\mev.

\section{Mass fit model}
\label{sec:fit}  

Using Monte Carlo simulations, We estimate resolutions which, in the mass region between 2000 and 2900~\mev, are similar for the three mass spectra and 
range from 1.0 to 4.5~\mev as a function of the mass. Since the widths of the resonances appearing in the three mass spectra are much larger than the experimental resolutions, resolution effects are neglected.

Binned $\chi^2$ fits to the three mass spectra are performed.
The \DTwentyFourSixty and \DTwentyFourHundred signal shapes in two-body decays are parameterized with a relativistic Breit-Wigner that includes the mass-dependent factors for a D-wave and S-wave decay, respectively.
The radius entering in the Blatt-Weisskopf~\cite{BW} form factor is fixed to 4~$\gev^{-1}$. 
Other resonances appearing in the mass spectra are described by Breit-Wigner lineshapes. 
All Breit-Wigner expressions are multiplied by two-body phase space.
The cross-feed lineshapes from \DTwentyFourTwenty and \DTwentyFourSixty appearing in the $\Dp \pim$ and $\Dz \pip$ mass spectra are described by a Breit-Wigner function fitted to the data.
The background $B(m)$ is described by an empirical shape~\cite{delAmoSanchez:2010vq} 
\begin{eqnarray}
B(m) = & P(m)e^{a_1m+a_2m^2} \ {\rm for} \ m<m_0, \nonumber\\
B(m) = & P(m)e^{b_0+b_1m+b_2m^2} \ {\rm for} \ m>m_0,
\end{eqnarray}
where $P(m)$ is the two-body phase space and $m_0$ is a free parameter. The two functions and their first derivatives are required to be continuous at $m_0$ and therefore the background model has four free parameters.

\begin{table}[b]
\caption{\small Definition of the categories selected by different ranges of \cthetah, and fraction of the total natural parity contribution.}
\label{tab:nat_unnat}
\begin{center}
\begin{tabular}{llc}
Category & Selection & natural parity fraction (\%)\cr
\hline
{\it Enhanced unnatural parity sample} & $|\cos \theta_{\rm H}| > 0.75 $ & \ 8.6 \cr
{\it Natural parity sample} &  $|\cos \theta_{\rm H}|<0.5$ & 68.8 \cr
{\it Unnatural parity sample} & $|\cos \theta_{\rm H}|>0.5$ & 31.2 \cr        
\hline    
\end{tabular}
\end{center}
\end{table}

\vspace{0.5cm}
\begin{boldmath}
\section{Fit to the $D^{*+}\pi^-$ mass spectrum}
\end{boldmath}

\label{sec:fit_dstarpi}

\begin{figure}[htb]
\centering
\includegraphics[height=2.0in]{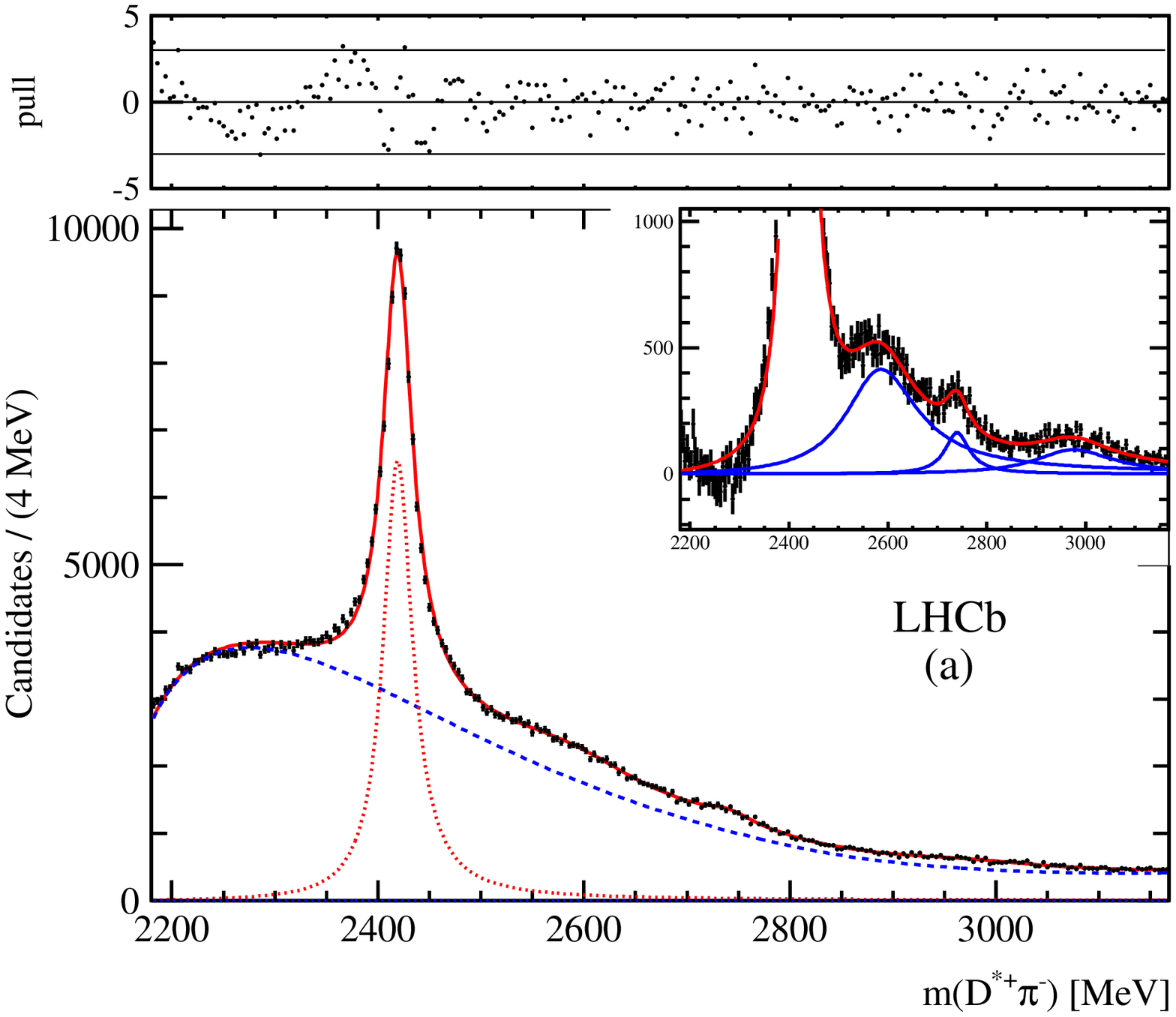}
\includegraphics[height=2.0in]{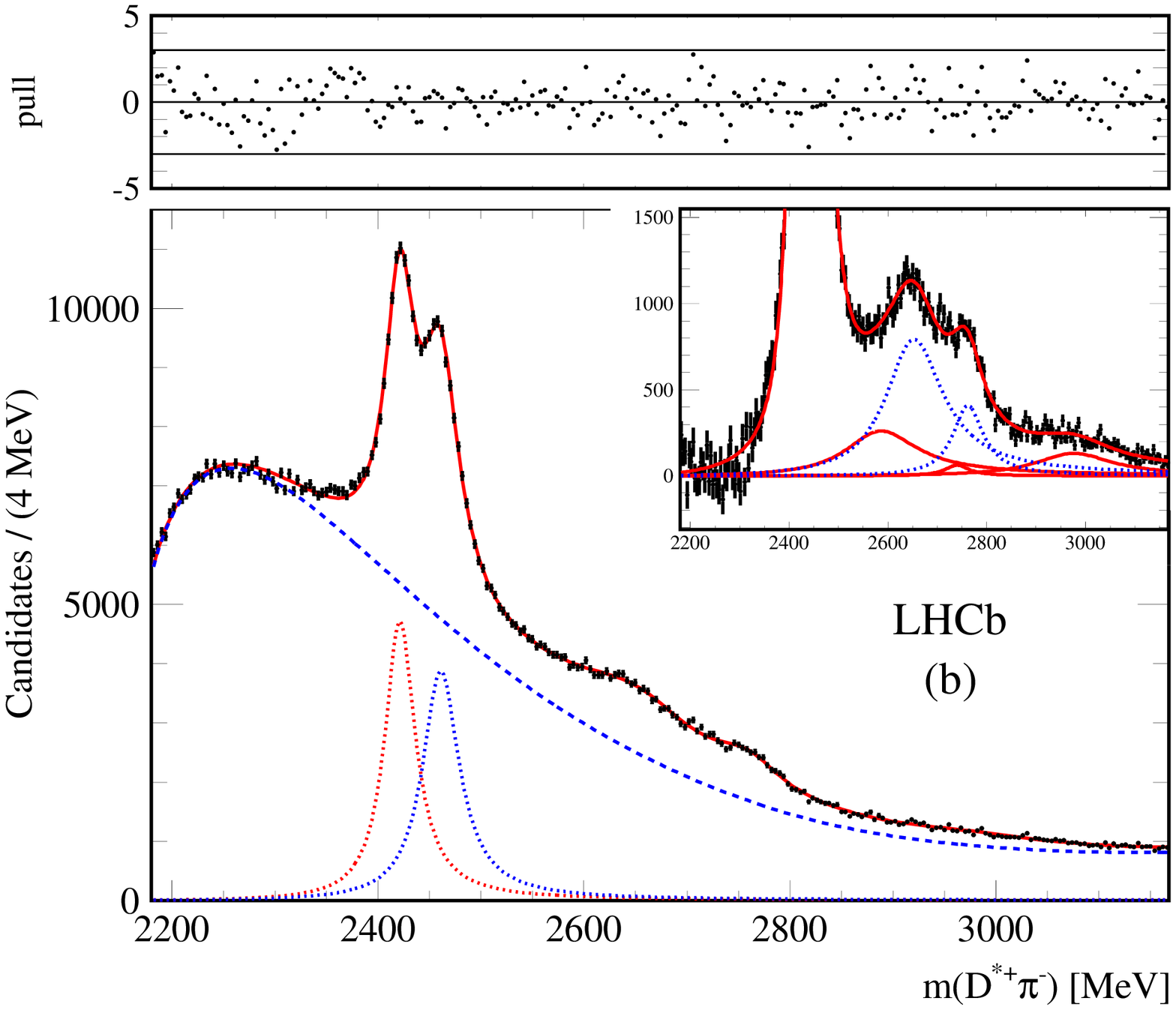}
  \caption{
    \small Fit to the $\Dstarp \pim$ mass spectrum, {\it enhanced unnatural parity sample}.  
The dashed (blue) line shows the fitted background, the dotted lines 
the \DTwentyFourTwentyNeutral (red) and  \DTwentyFourSixtyNeutral (blue) contributions. The inset displays the $\Dstarp \pim$ mass spectrum after subtracting the fitted background. The full line curves (red) show the contributions from \DTwentyFiveFiftyNeutral, \DTwentySevenFiftyNeutral, and \DThreeU. The dotted (blue) lines display the \DTwentySixHundredNeutral and \DTwentySevenSixtyNeutral contributions. The top window shows the pull distribution where the horizontal lines indicate $\pm 3 \sigma$. The pull is defined as $(N_{\rm data} - N_{\rm fit})/\sqrt{N_{\rm data}}$.
    }
  \label{fig:fig4}
\end{figure}

Due to the three-body decay and the availability of the helicity angle information, the fit to the $\Dstarp \pim$ mass spectrum allows a spin analysis of the produced resonances and a 
separation of the different spin-parity components. We define the helicity angle \mthetah as the angle between the $\pim$ and the $\pip$ from the $\Dstarp$ decay, in the rest frame of the $\Dstarp \pim$ system. 
Full detector simulations are used to measure the efficiency as a function of \mthetah, which is found to be uniform. 

It is expected that the angular distributions are proportional to $\sin^2\mthetah$ for natural parity resonances and proportional to $1+h\cos^2\mthetah$ for unnatural parity resonances, where $h>0$ is a free parameter. 
The $\Dstar \pi$ decay of a $J^P=0^+$ resonance is forbidden. 
Therefore candidates
selected in different ranges of \cthetah can enhance or suppress the different spin-parity contributions. We separate the $\Dstarp \pim$ data into three different categories, summarized in Table~\ref{tab:nat_unnat}. 

The data and fit for the $\Dstarp \pim$  {\it enhanced unnatural parity sample} are shown in Fig.~\ref{fig:fig4}(a) and the resulting fit
parameters are summarized in Table~\ref{tab:fits}. 
The mass spectrum is dominated by the presence of the unnatural parity \DTwentyFourTwentyNeutral resonance. 
The fitted natural parity \DTwentyFourSixtyNeutral contribution is consistent with zero, as expected. 
To obtain a good fit to the mass spectrum, three further resonances are needed. We label them \DTwentyFiveFiftyNeutral, \DTwentySevenFiftyNeutral, and \DThreeU.
The presence of these states in this sample indicates unnatural parity assignments.

The masses and widths of the unnatural parity resonances are fixed in the fit to the {\it natural parity sample}.
The fit is shown in Fig.~\ref{fig:fig4}(b) and the obtained resonance 
parameters are summarized in Table~\ref{tab:fits}. 
The mass spectrum shows that the unnatural parity resonance \DTwentyFourTwentyNeutral is suppressed with respect to that observed in the
 {\it enhanced unnatural parity sample}. 
There is a strong contribution of the natural parity \DTwentyFourSixtyNeutral resonance and
contributions from the \DTwentyFiveFiftyNeutral, \DTwentySevenFiftyNeutral and \DThreeU states.
To obtain a good fit, two additional resonances are needed, which we label \DTwentySixHundredNeutral and \DTwentySevenSixtyNeutral.

Table~\ref{tab:fits} summarizes the measured resonance parameters and yields. 
The significances are computed as $\sqrt{\Delta \chi^2}$ where $\Delta \chi^2$ is 
the difference between the $\chi^2$ values when a resonance is included or excluded from the fit while all the other resonances parameters are allowed to vary.  All the statistical significances are well above 5$\sigma$.

\begin{table}
{\scriptsize
\caption{\small Resonance parameters, yields and statistical significances. The first uncertainty is statistical, the second systematic.}
\label{tab:fits}
\begin{center}
\begin{tabular}{c | c | r@{}c@{}l | r@{}c@{}l | r@{}c@{}l | c}
Resonance & Final state & \multicolumn{3}{c|}{Mass (MeV)} & \multicolumn{3}{c|}{Width (MeV)} & \multicolumn{3}{c|}{Yields $\times 10^3$} & Significance \cr
\hline
\DTwentyFourTwentyNeutral & $\Dstarp \pim$ & 2419.6 $\pm$ & \, 0.1 \, &  $\pm$ 0.7 &  35.2 $\pm$ & \, 0.4 \, &  $\pm$ 0.9  &  210.2 $\pm$ & \, 1.9 \, &  $\pm$ 0.7 &   \,   \\
\DTwentyFourSixtyNeutral  & $\Dstarp \pim$ & 2460.4 $\pm$ & \, 0.4 \, &  $\pm$ 1.2 &  43.2 $\pm$ & \, 1.2 \, &  $\pm$ 3.0  &   81.9 $\pm$ & \, 1.2 \, &  $\pm$ 0.9 &   \,    \\
\DTwentySixHundredNeutral & $\Dstarp \pim$ & 2649.2 $\pm$ & \, 3.5 \, &  $\pm$ 3.5 & 140.2 $\pm$ & \, 17.1 \,&  $\pm$ 18.6 &   50.7 $\pm$ & \, 2.2 \, &  $\pm$ 2.3 &  24.5     \\
\DTwentySevenSixtyNeutral & $\Dstarp \pim$ & 2761.1 $\pm$ & \, 5.1 \, &  $\pm$ 6.5 &  74.4 $\pm$ & \, 3.4 \, &  $\pm$ 37.0 &   14.4 $\pm$ & \, 1.7 \, &  $\pm$ 1.7 &  10.2     \\
\DTwentyFiveFiftyNeutral  & $\Dstarp \pim$ & 2579.5 $\pm$ & \, 3.4 \, &  $\pm$ 5.5 & 177.5 $\pm$ & \, 17.8 \, & $\pm$ 46.0 &   60.3 $\pm$ & \, 3.1 \, &  $\pm$ 3.4 &  18.8   \\
\DTwentySevenFiftyNeutral & $\Dstarp \pim$ & 2737.0 $\pm$ & \, 3.5 \, &  $\pm$11.2 &  73.2 $\pm$ & \, 13.4 \, & $\pm$ 25.0 &    7.7 $\pm$ & \, 1.1 \, &  $\pm$ 1.2 &  \ 7.2     \\
\DThreeU                  & $\Dstarp \pim$ & 2971.8 $\pm$ & \, 8.7 \, &            & 188.1 $\pm$ & \, 44.8 \, &            &    9.5 $\pm$ & \, 1.1 \, &             &  \ 9.0     \\
\hline
\DTwentyFourSixtyNeutral  & $\Dp \pim$     & 2460.4 $\pm$ & \, 0.1 \, & $\pm$ 0.1  &  45.6 $\pm$ & \,  0.4 \, & $\pm$ 1.1 &   675.0 $\pm$ & \, 9.0 \, &  $\pm$ 1.3 &                   \\
\DTwentySevenSixtyNeutral & $\Dp \pim$     & 2760.1 $\pm$ & \, 1.1 \, & $\pm$ 3.7  &  74.4 $\pm$ & \,  3.4 \, & $\pm$19.1 &    55.8 $\pm$ & \, 1.3 \, &  $\pm$ 10.0 &   17.3     \\
\DThreeNeutral            & $\Dp \pim$     & 3008.1 $\pm$ & \, 4.0 \, &            & 110.5 $\pm$ & \, 11.5 \, &            &   17.6 $\pm$  & \,1.1 \, &             &   21.2   \\
\hline 
\DTwentyFourSixtyCharged  & $\Dz \pip$     & 2463.1 $\pm$ & \, 0.2 \, & $\pm$ 0.6  &  48.6 $\pm$ & \,  1.3 \, & $\pm$ 1.9 &   341.6 $\pm$ & \,22.0 \, &  $\pm$ 2.0 &                   \\
\DTwentySevenSixtyCharged & $\Dz \pip$     & 2771.7 $\pm$ & \, 1.7 \, & $\pm$ 3.8  &  66.7 $\pm$ & \,  6.6 \, & $\pm$10.5 &    20.1 $\pm$ & \, 2.2 \, &  $\pm$ 1.0 &   18.8      \\
\DThreeCharged            & $\Dz \pip$     & 3008.1       & \, (fixed) \,   &        & 110.5       & \, (fixed) \, &           &    7.6 $\pm$  & \, 1.2 \, &             & \  6.6    \\
\hline
\end{tabular}
\end{center}
}
\end{table}

\vspace{0.5cm}
\begin{boldmath}
\section{Spin-parity analysis of the \DstarPi system}
\end{boldmath}
\label{sec:spin}

In order to obtain information on the spin-parity assignment of the states observed in the  $\Dstarp \pim$ mass spectrum,
the data are subdivided into ten equally spaced bins in $\cthetah$. The ten mass spectra are then fitted
with the model described above with fixed resonance parameters to obtain the yields as functions of $\cthetah$ for each resonance.

\begin{figure}[htb]
\centering
\includegraphics[height=1.4in]{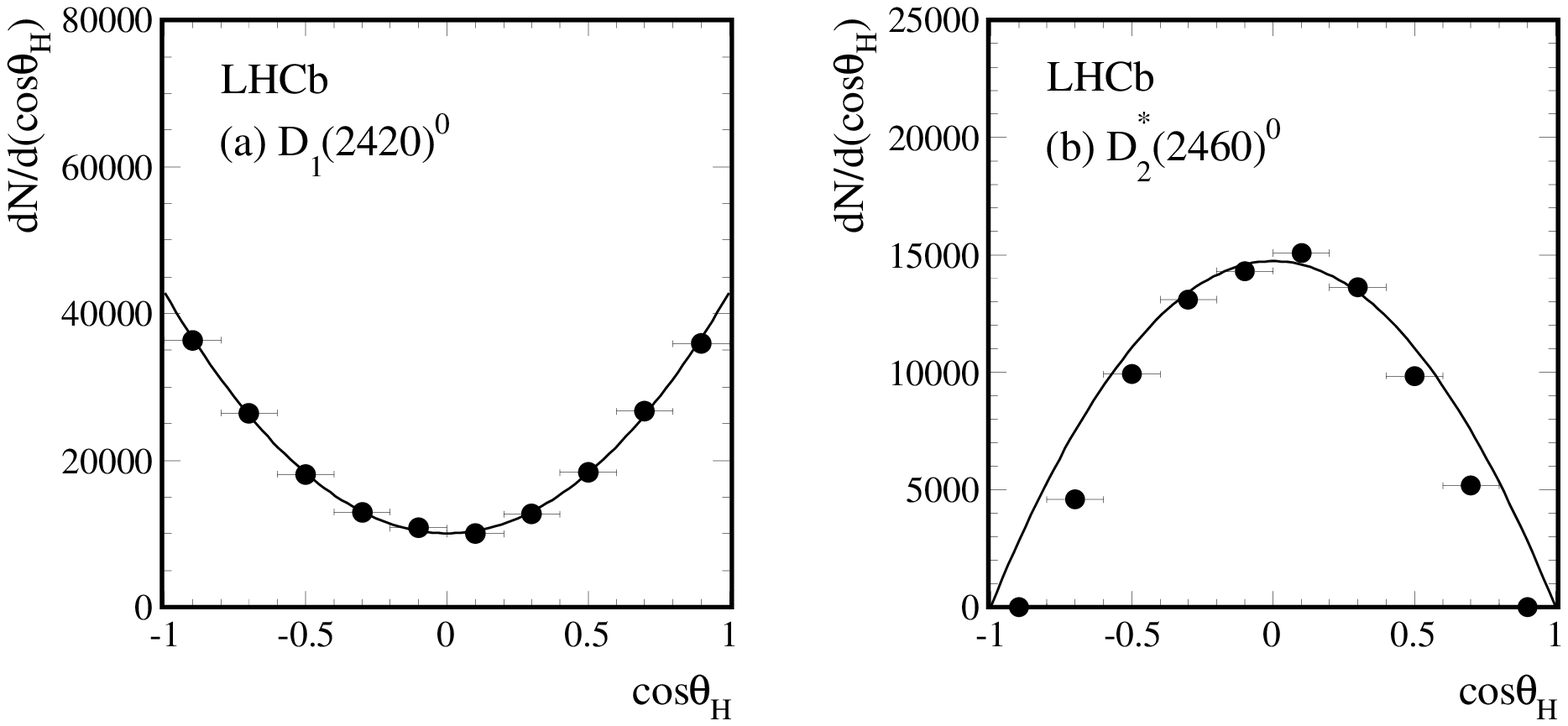}
\includegraphics[height=1.4in]{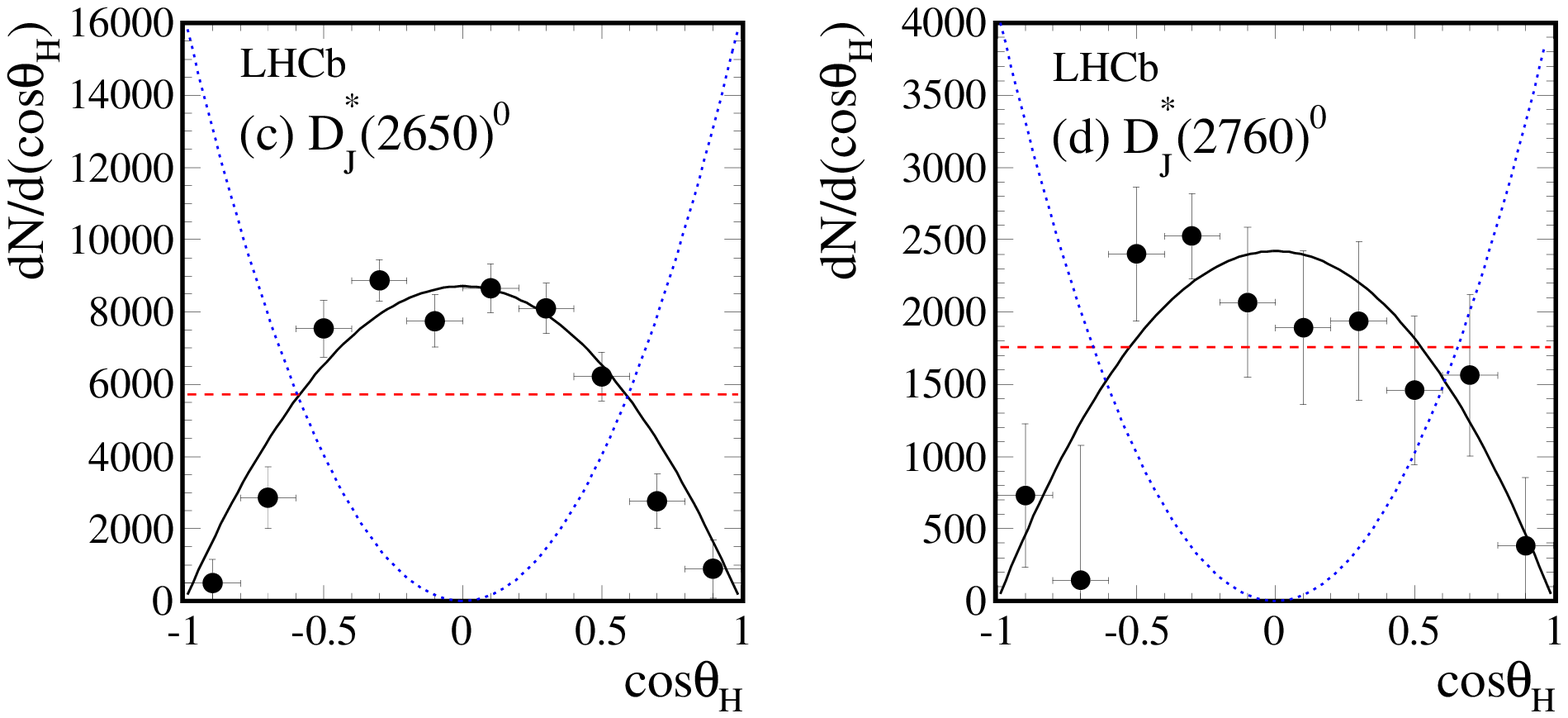}
  \caption{
    \small Distributions of (a) \DTwentyFourTwentyNeutral, (b) \DTwentyFourSixtyNeutral, (c) \DTwentySixHundredNeutral and (d) \DTwentySevenSixtyNeutral candidates as functions of the 
helicity angle \cthetah. The distributions are fitted with 
natural parity (black continuous), unnatural parity (red, dashed) and $J^P=0^-$ (blue, dotted) functions.
    }
  \label{fig:fig8}
\end{figure}

The resulting distributions for \DTwentyFourTwentyNeutral and \DTwentyFourSixtyNeutral are shown in Fig.~\ref{fig:fig8}(a)-(b). A good description of the data is obtained in terms of the expected angular distributions for $J^P=1^+$ and $J^P=2^+$ resonances.

Figure~\ref{fig:fig8}(c)-(d) shows the resulting distributions for the \DTwentySixHundredNeutral and \DTwentySevenSixtyNeutral states. In this case we compare the distributions with expectations from natural parity, unnatural parity and $J^P=0^-$. In the case of unnatural parity, the $h$ parameter, in $ 1+h\cos^2\theta_{\rm H}$,  is constrained to be positive and therefore the fit gives $h=0$. In both cases, the distributions are best fitted by the 
natural parity hypothesis.

\begin{figure}[htb]
\centering
\includegraphics[height=1.5in]{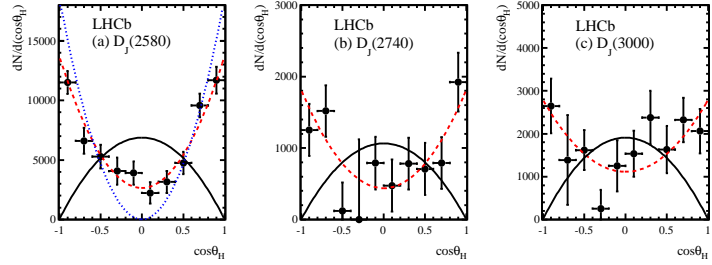}
  \caption{
    \small Distributions of (a) \DTwentyFiveFiftyNeutral, (b) \DTwentySevenFiftyNeutral and (c) \DThreeU candidates as functions of the 
helicity angle \cthetah. The distributions are fitted with 
natural parity (black continuous) and unnatural parity (red, dashed) functions. In (a) the $J^P=0^-$ (blue, dotted) hypothesis is also tested. 
    }
  \label{fig:fig10}
\end{figure}

Figure~\ref{fig:fig10} shows the angular distributions for the \DTwentyFiveFiftyNeutral, \DTwentySevenFiftyNeutral and \DThreeU states. The distributions are fitted with 
natural parity and unnatural parity. The $J^P=0^-$ hypothesis is also considered for \DTwentyFiveFiftyNeutral. In all cases unnatural parity is preferred over
a natural parity assignment.

\vspace{0.5cm}
\begin{boldmath}
\section{Fit to the \DPi and \DzPi mass spectra}
\end{boldmath}
\label{sec:fit_dpi}

The $\Dp \pim$ and $\Dz \pip$ mass spectra consist of natural parity resonances.
However these final states are affected by cross-feed from all the resonances that decay to the $\Dstar \pi$ final state. Figures~\ref{fig:fig3}(a)-(b) show 
(in the mass region around 2300 MeV) cross-feed contributions from 
\DTwentyFourTwenty and \DTwentyFourSixty decays. However we also expect (in the mass region between 2400 and 2600 MeV) the presence of structures originating from the complex resonance structure present in the $\Dstar \pi$ mass spectrum in the mass region between 2500 and 2800~\mev.

\begin{figure}[htb]
\centering
\includegraphics[height=2.0in]{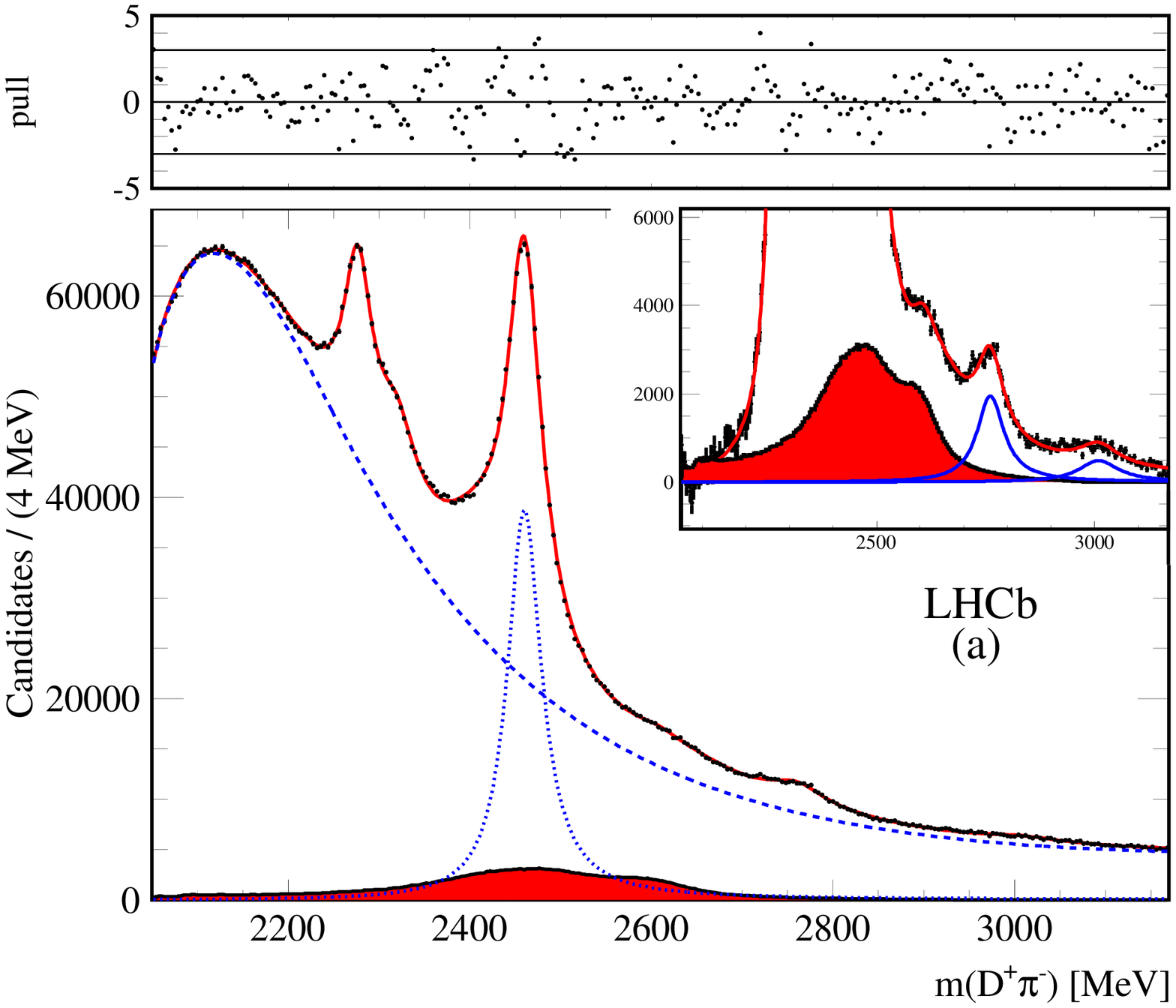}
\includegraphics[height=2.0in]{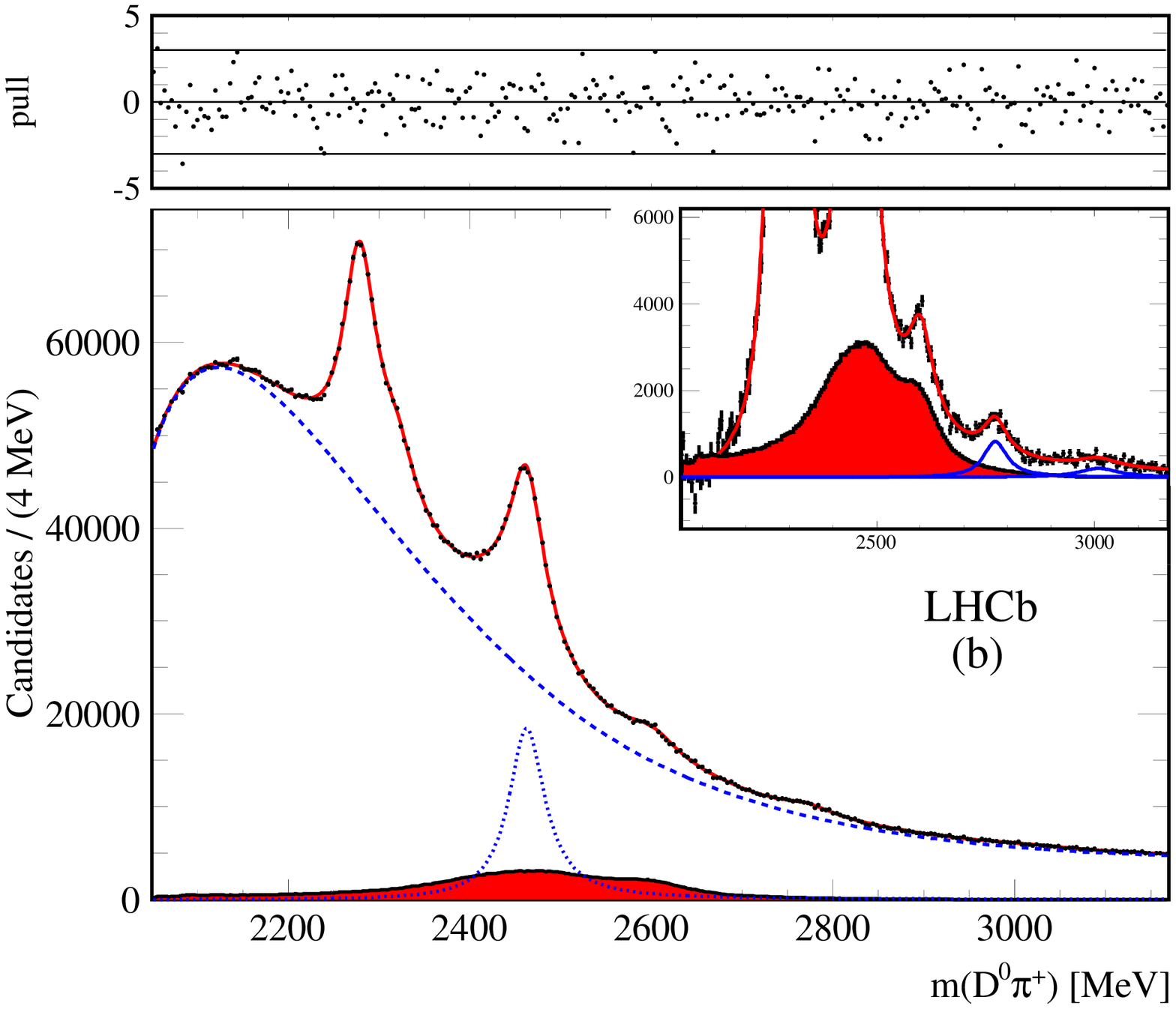}
  \caption{
    \small (a) Fit to the $\Dp \pim$ mass spectrum and (b) to the $\Dz \pip$ mass spectrum. The filled histogram (in red) shows the estimated cross-feeds from the high mass $\Dstar \pi$ resonances. 
    }
  \label{fig:fig11}
\end{figure}

To obtain an estimate of the lineshape and size of the cross-feed, we normalize the $\Dstarp \pim$ mass spectrum to the $\Dp \pim$ mass spectrum using 
the sum of the \DTwentyFourTwentyNeutral and \DTwentyFourSixtyNeutral yields in the $\Dstarp \pim$ mass spectrum and the sum of the cross-feed in the 
$\Dp \pim$ mass spectrum. To obtain the expected lineshape of the cross-feed in the $\Dp \pim$ final state, we perform a study based on a generator level simulation.
We generate \DTwentySixHundredNeutral, \DTwentySevenSixtyNeutral, \DTwentyFiveFiftyNeutral and \DTwentySevenFiftyNeutral decays according to the chain described in Eq.~(\ref{dp_pim}).
We then compute the resulting $\Dp \pim$ mass spectra and normalize each contribution to the measured yields. The overall resulting structures
are then properly scaled and superimposed on the $\Dp \pim$ mass spectrum shown in Fig.~\ref{fig:fig11}(a). A similar method is used for the $\Dz \pip$ final state and the resulting contribution is superimposed on the $\Dz \pip$ mass spectrum shown in Fig.~\ref{fig:fig11}(b). To obtain good quality fits we add broad structures around 3000~\mev, which we label \DThreeNeutral and \DThreeCharged.

The fits to the $\Dp \pim$ and $\Dz \pip$ mass spectra are shown in Fig.~\ref{fig:fig11}(a) and Fig.~\ref{fig:fig11}(b), respectively.
Several cross-checks are performed to test the stability of the fits and their correct statistical behaviour.
We first repeat all the fits, including the spin-parity analysis, lowering the \pt requirement from 7.5 to 7.0 GeV. We find that all the resonance parameters vary within their statistical uncertainties and that the spin-parity assignments are not affected by this selection. 
Then we perform fits using random variations of the histogram contents and background parameters.
The various estimated systematic uncertainties are added in quadrature.

\vspace{1cm}

\section{Precision measurement of $D$ meson mass differences}
\label{sec:dmass}

Using three- and four-body decays of $D$ mesons produced in semileptonic
$b$-hadron decays, precision measurements of $D$ meson mass differences are
made together with a measurement of the $D^{0}$ mass~\cite{Aaij:2013uaa}. 
The selection uses only well reconstructed charged
particles that traverse the entire tracking system.
Further background suppression is achieved by exploiting
the fact that the products of heavy flavour decays have a large
distance of closest approach (`impact parameter') with respect to the $pp$ interaction vertex in which they were
produced. The impact parameter $\chi^2$ with respect to any
primary vertex is required to be larger than nine.
  
Charged particles are combined to form $\D^{0}
\rightarrow K^{+} K^{-} \pi^{+} \pi^{-} $,  $\D^{0} \rightarrow K^{+}
K^{-} K^{-} \pi^{+} $ and  $\D^{+}_{(s)} \rightarrow K^{+} K^{-}
\pi^{+} $  candidates.
To eliminate kinematic reflections due to 
misidentified pions, the invariant mass of at least one kaon pair is required to be within $ \pm 12~\mevcc$ of the 
nominal value of the $\Pphi$~meson mass. 
Each candidate $D$ meson is combined with a
well-identified muon that is displaced from the $pp$ interaction vertex to form a $B$ candidate, requiring the muon and the $D$ candidate to originate from a common point. 

\begin{figure}[htb]
\centering
\includegraphics[height=1.5in]{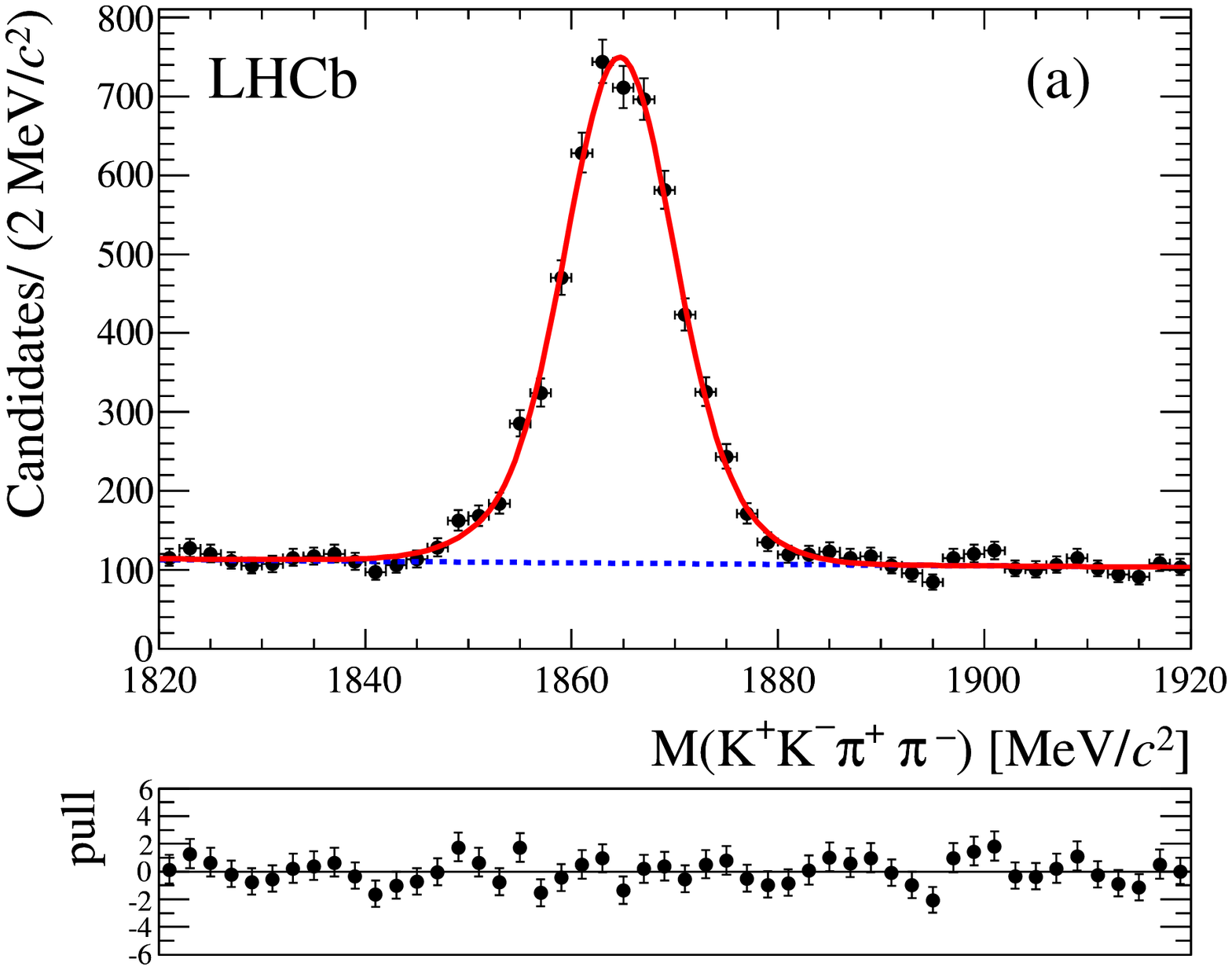}
\includegraphics[height=1.5in]{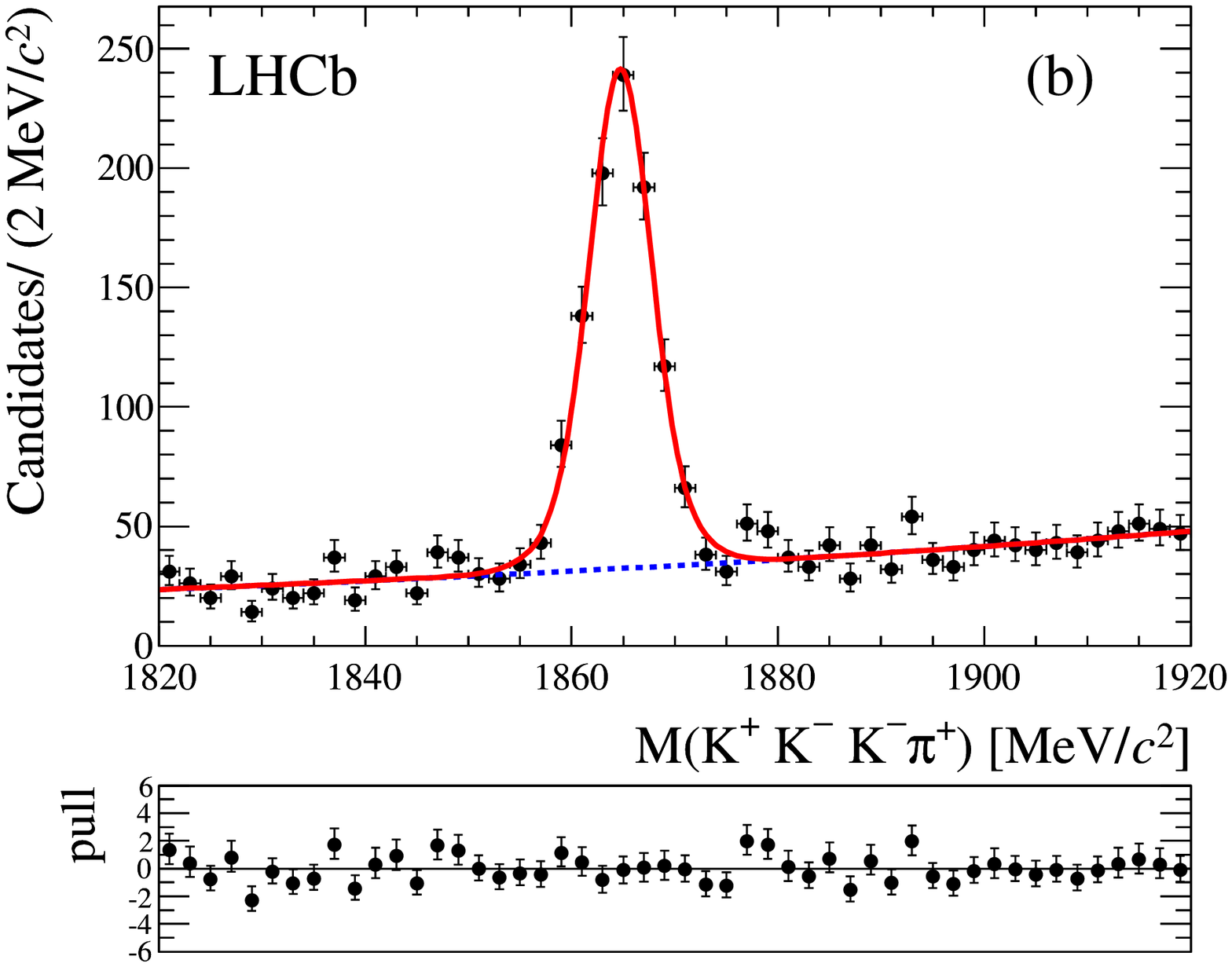}
\includegraphics[height=1.5in]{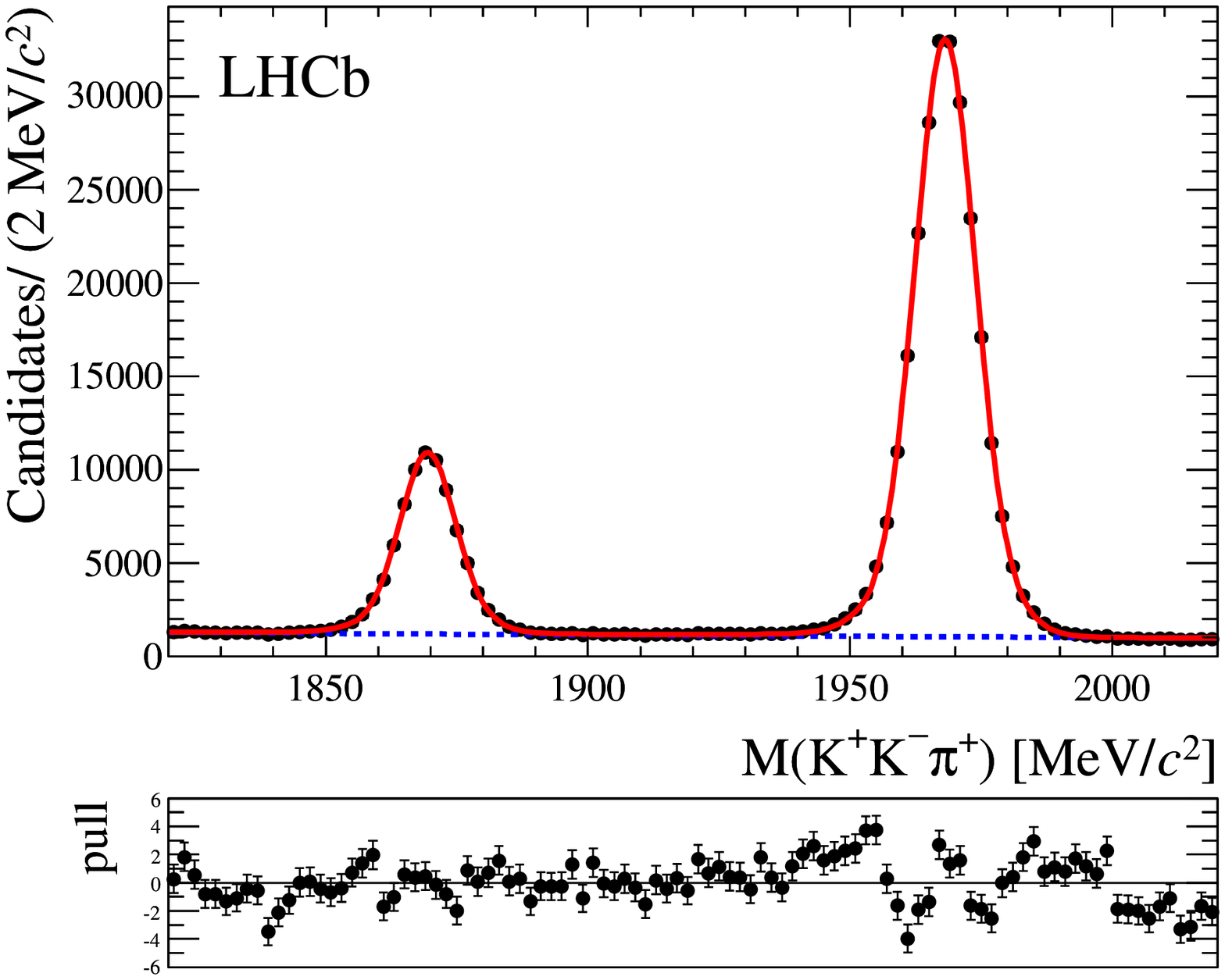}
  \caption{
    \small Invariant mass distributions for  
the (a) $K^{+} K^{-} \pi^{+} \pi^{-} $  
and (b) $K^{+} K^{-} K^{-} \pi^{+}$ final states. Invariant mass distribution for  
the $K^{+} K^{-} \pi^{+}$ final state.
    }
  \label{fig:d0Masses}
\end{figure}

The $D$ meson masses are determined by performing extended unbinned maximum
likelihood fits to the invariant mass distributions. In these fits the
background is modelled by an exponential function and the signal by
the sum of a Crystal Ball \cite{Skwarnicki:1986xj} and a Gaussian function. The Crystal Ball
component accounts for the presence of the QED radiative tail.
The fits for the $D^0$ decay modes and the $K^{+} K^{-} \pi^+$ final state
are shown in Fig.~\ref{fig:d0Masses}. 

The resulting
values of the $D^+$ and $D^+_s$ masses 
are in agreement with the current world averages. These modes
have relatively large $Q$-values and consequently the 
systematic uncertainty due to the knowledge of the momentum scale 
is at the level of $0.3\,\mevcc$. Hence, it is chosen not to
quote these values as measurements. Similarly, the systematic
uncertainty due to the momentum
scale for the $\D^{0} \rightarrow K^{+} K^{-} \pi^{+} \pi^{-}$
mode is estimated to be $0.2~\mevcc$ and the measured mass
in this mode is not used in the $\D^0$ mass determination. 

We obtain
\begin{center}
\begin{tabular}{l @{$~=~$}l@{$\,\,\pm\,$}l@{\,(stat) $\pm\,$}l@{\,(syst) MeV/$c^2$}l}
$M(D^0)$  & 1864.75 & 0.15 & 0.11 & , \\
$M(D^{+})$ $-$ $M(D^{0})$   & \phantom{111}4.76 & 0.12 & 0.07 & , \\
$M(D^{+}_s)$ $-$ $M(D^{+})$   & \phantom{11}98.68 & 0.03 & 0.04 & , \\
\end{tabular}
\end{center}
where dominant systematic uncertainty is related to the knowledge of
the momentum scale.
 
The measurements presented here, together with those given in
Ref.~\cite{Beringer:1900zz} for the $D^+$ and $D^0$ mass, and the mass
differences $M(D^+) -M(D^0)$, $M(D^{+}_s) - M(D^{+})$ 
can be used to determine a more precise value of the $D^+_s$ mass
\begin{equation}
M(D^{+}_s)  =  1968.19 \pm 0.20 \pm 0.14 \pm 0.08 \mevcc, \nonumber
\end{equation}
where the first uncertainty is the quadratic sum of the statistical
and uncorrelated systematic uncertainty, the second is due to the
momentum scale and the third due to the energy loss. This value is 
consistent with, but more precise than, that obtained from the 
fit to open charm mass data, $M(D^{+}_s) =  1968.49 \pm 0.32~\mevcc$ \cite{Beringer:1900zz}. 
\Acknowledgements

\noindent We express our gratitude to our colleagues in the CERN
accelerator departments for the excellent performance of the LHC. We
thank the technical and administrative staff at the LHCb
institutes. We acknowledge support from CERN and from the national
agencies: CAPES, CNPq, FAPERJ and FINEP (Brazil); NSFC (China);
CNRS/IN2P3 and Region Auvergne (France); BMBF, DFG, HGF and MPG
(Germany); SFI (Ireland); INFN (Italy); FOM and NWO (The Netherlands);
SCSR (Poland); MEN/IFA (Romania); MinES, Rosatom, RFBR and NRC
``Kurchatov Institute'' (Russia); MinECo, XuntaGal and GENCAT (Spain);
SNSF and SER (Switzerland); NAS Ukraine (Ukraine); STFC (United
Kingdom); NSF (USA). We also acknowledge the support received from the
ERC under FP7. The Tier1 computing centres are supported by IN2P3
(France), KIT and BMBF (Germany), INFN (Italy), NWO and SURF (The
Netherlands), PIC (Spain), GridPP (United Kingdom). We are thankful
for the computing resources put at our disposal by Yandex LLC
(Russia), as well as to the communities behind the multiple open
source software packages that we depend on.

\end{document}